\documentclass{PoS}

\usepackage{cite}

\newcommand{\qqbar}{q\overline{q}}
\newcommand{\mumu}{\mu^+\mu^-}
\newcommand{\pipi}{\pi^+\pi^-}
\newcommand{\KK}{K^+K^-}
\newcommand{\mee}{e^+e^-}
\def\GeVM{~${\rm GeV}/c^2$}

\def\GeVp{~${\rm GeV}/c$}
\def\GeVE{~${\rm GeV}$}

\def\intl{\int\limits}

\title{Measurement of hadronic cross sections at BaBar with ISR and implications for the muon (g-2)}

\ShortTitle{Measurement of hadronic cross sections at BaBar with ISR and implications for the muon (g-2)}

\author{\speaker{Bogdan Malaescu}
        \thanks{The author would like to thank the organizers for the invitation to attend the conference.}\\
        On behalf of the BABAR collaboration \\
        LAL Orsay, now at LPNHE Paris\\
        E-mail: \email{malaescu@in2p3.fr}}

\abstract{The ISR method has been largely exploited by the BABAR experiment, for measuring numerous channels of the cross section $\mee \to {\rm hadrons}$.
For the $\pipi(\gamma)$ and $\KK(\gamma)$ channels, BABAR has pioneered the method based on the ratio between the hadronic mass spectra and the $\mumu(\gamma)$ one.
This method allows to cancel many systematic uncertainties in the ratio, hence the precise measured cross sections.
Many multihadronic channels have also been studied using the ISR method, and cross sections have been published.
These experimental results have also been exploited for phenomenological studies, like the determination of the hadronic contribution to the anomalous magnetic moment of the muon $(g-2)_\mu$.}

\FullConference{International Conference on the Structure and the Interactions of the Photon including the 20th International Workshop on Photon-Photon Collisions and the International Workshop on High Energy Photon Linear Colliders\\
                 20 - 24 May 2013\\
                 Paris, France}

\begin{document}

\section{Introduction}

Precise measurements of the $\mee\to {\rm hadrons}$ cross-section are needed for various phenomenological studies, which motivated the BABAR extensive program for measuring them~\cite{Aubert:2009ad,Lees:2012cj,Lees:2013gzt,Lees:2013uta,:2013xe,Aubert:2005cb,Lees:2012cr,Lees:2011zi,Aubert:2007ym,Aubert:2007ur,Aubert:2007uf,Aubert:2007ef,Aubert:2006bu,Aubert:2006xw,Aubert:2006jq,Aubert:2005eg,Aubert:2004kj}.
In particular, they are used to evaluate dispersion integrals for calculations of the hadronic vacuum polarization~(VP). 
A well known example is the hadronic contribution to the muon magnetic moment anomaly~($a_\mu^{had}$), requiring data in the low mass region.
It is dominated by the process $\mee\to\pipi(\gamma)$ which provides $73\%$ of the contribution, bringing also the dominant contribution to the uncertainty.

Recent measurements of the $\pi\pi$ cross section, previous to the BABAR publications, have a systematic precision of $0.8\%$ for CMD2~\cite{cmd-2} and 
$1.5\%$ for SND~\cite{snd}.
These two measurements are in good agreement. 
The first measurement using the ISR method~\cite{isr}, done by KLOE~\cite{kloe04}, had a quoted systematic precision of $1.3\%$.
However, some significant deviation in shape was observed when comparing to the Novosibirsk data. 
The KLOE data were reanalysed~\cite{kloe08} and the agreement with the Novosibirsk data is improved.
For the updated measurement, a systematic uncertainty of $0.9\%$ is quoted.

When previous $\mee$ data~\cite{cmd-2,snd,kloe08} are used, the comparison of the theoretical and measured~\cite{bnl} values of $a_\mu$ 
shows a discrepancy of about three standard deviations.
This is a possible hint at new physics. 
When using an approach based on hadronic $\tau$ decay data, corrected for isospin-breaking effects, a smaller difference is observed~\cite{newtau}. 

In these proceedings we present the BABAR $2\pi(\gamma)$ result, published in \cite{Aubert:2009ad,Lees:2012cj}.
This study yielded a measurement of the contribution of the $2\pi$ channel to ($a_\mu^{had}$) with a precision better than $1\%$.
This implies a control of systematic uncertainties at the $10^{-3}$ level. 
We also discuss the $2K(\gamma)$ measurement, presented as preliminary result at the conference, now published in \cite{Lees:2013gzt}, as well as several
BABAR measurements of multihadronic channels.

\section{The BABAR ISR $\pipi$, $\KK$ and $\mumu$ analyses} 

The measurements of the $\pi\pi$ and $\rm KK$ cross sections presented here are performed using the ISR method~\cite{isr} for $\mee$ annihilation
events collected with the BABAR detector, at a center-of-mass energy $\sqrt{s}$ near $10.58$\GeVE.
We consider events $\mee \to X\gamma_{ISR}$, where $X$ can correspond to any final state, and the ISR~photon is emitted by the $e^+$ or $e^-$.
The $\mee \to \pi\pi(\gamma_{FSR})$ and $\mee \to \rm KK(\gamma_{FSR})$ cross sections are obtained as a function of the invariant mass of the final state $\sqrt{s'}$.
The advantage of the ISR method~(compared to an energy scan) is that all the mass spectrum is covered at once~(from threshold to $3~(5)$\GeVE~ for $\pi\pi~(\rm KK)$ in BABAR) with the same detector conditions and analysis.

In this BABAR study the $\pipi\gamma_{ISR}(\gamma_{FSR})$, $\KK\gamma_{ISR}(\gamma_{FSR})$ and $\mumu\gamma_{ISR}(\gamma_{FSR})$ spectra are measured.
These are the first NLO measurements, a possible additional radiation being taken into account in the analysis, instead of being corrected
a posteriori (as done by other experiments).
The measured muon spectrum is compared with the NLO QED prediction.
This represents an important cross check of the analysis, called the QED test.
The cross section for the process $\mee\to X$ is related to the $\sqrt{s'}$ spectrum of $\mee\to X\gamma$ events through
\begin{equation}
\label{def-lumi}
  \frac {\mathrm{d}N_{X\gamma}}{\mathrm{d}\sqrt{s'}}~=~\frac {\mathrm{d}L_{ISR}^{eff}}{\mathrm{d}\sqrt{s'}}~
    \varepsilon_{X\gamma}(\sqrt{s'})~\sigma_{X}^0(\sqrt{s'})~,
\end{equation}
where $\sigma_X^0$ is the bare cross section~(excluding VP),
and $\varepsilon_{X\gamma}$ is the detection efficiency~(acceptance) determined by simulation with corrections obtained from data.
The effective ISR luminosity $\rm{L}^{\rm eff}_{\rm ISR}$ is derived using the muon spectrum.
The contribution of leading order FSR for muons~(smaller than $1\%$, below $1$\GeVE) is corrected for, while additional FSR photons are measured. 
The $\pi\pi(\gamma_{FSR})$ and $\rm KK(\gamma_{FSR})$ cross sections are obtained from the ratio of the corresponding hadronic spectra and $\rm{L}^{\rm eff}_{\rm ISR}$.
The $\mee$ luminosity, additional ISR effects, vacuum polarization and ISR photon efficiency cancel in the ratio, hence the strong reduction of the systematic uncertainty. 

This analysis is based on $232~{\rm fb}^{-1}$ of data recorded at the PEP-II asymmetric-energy $\mee$ storage rings, with the BABAR detector~\cite{detector}.
The energy and direction of photons are measured in the CsI(Tl) electromagnetic calorimeter (EMC).
Charged-particle tracks are measured with a five-layer double-sided silicon vertex tracker (SVT) together with a 40-layer drift chamber (DCH) inside a 1.5 T superconducting solenoid magnet.
The identification of charged-particles (PID) uses ionization loss ${\rm d}E/{\rm d}x$ in the SVT and DCH, the Cherenkov radiation detected in a ring-imaging device (DIRC), and the shower deposit in the EMC ($E_{cal}$) and in the instrumented flux return (IFR) of the magnet.

The selection of two-body ISR events is done requiring a photon with $E_\gamma^*>3$\GeVE~ and laboratory polar angle in the range $0.35-2.4~{\rm rad}$, 
as well as exactly two tracks of opposite charge, each with momentum $p>1$\GeVp~ and within the angular range $0.40-2.45~{\rm rad}$. 
The events with one single charged track are also recorded and used for in-situ efficiency measurements. 

The simulation of signal and background ISR processes is done with Monte Carlo (MC) event generators based on Ref.~\cite{eva}.
The structure function method~\cite{struct-fct} is used to generate additional ISR photons, while {\small PHOTOS}~\cite{photos} is used for additional FSR photons.
The simulation of the BABAR detector is done with {\small GEANT4}~\cite{geant}. 

Background events from $\mee\to\qqbar$ ($q=u,d,s,c$) are generated using {\small JETSET}~\cite{jetset}. 
They are due to events with low-multiplicities and an energetic $\gamma$ from a $\pi^0$ mistaken as the ISR photon candidate. 
The data/MC comparison of the $\pi^0$ yield (obtained by pairing the ISR photon with other photons in the event) is used to normalize this rate from {\small JETSET}. 
The contributions from $\mee\to\pi^+\pi^-\pi^0\gamma$ and $\mee\to\pi^+\pi^-2\pi^0\gamma$ ISR backgrounds are dominant for the $\pipi$ channel.
In the $\KK(\gamma)\gamma_{\rm ISR}$ sample backgrounds stem mainly from
other ISR events: $\pi^+\pi^-\gamma$, $\mu^+\mu^-\gamma$, $\KK\eta\gamma$,
$\KK\pi^0\gamma$, $\pi^+\pi^-\pi^0\gamma$, $\pi^+\pi^-2\pi^0\gamma$,
$p\bar{p}\gamma$, and $K_SK_L\gamma$.
The background level from the $3\pi$ ISR process, is calibrated using $\omega$ and $\phi$ signals, following an approach similar to that for $\qqbar$.
The MC estimate for the $2\pi 2\pi^0\gamma$ process is used, with an assigned systematic uncertainty of $10\%$.
For the $\mu\mu$ spectrum background contributions are found to be negligible.

The simulation is used to compute the acceptance and mass-dependent efficiencies for trigger, reconstruction, PID, and event selection.
Specific studies, as described below, are use to determine the ratios of data and MC efficiencies,
which are then applied as mass-dependent corrections to the MC efficiency.
They amount to at most a few percent and are known to a few permil level or better.

Tracking and PID efficiencies are determined with a tag-and-probe method, taking advantage of pair production. 
Two-prong ISR candidates are selected for tracking studies, on the basis of the ISR photon and one track.  
The expected parameters of the second track are derived with a kinematic fit. 
The track reconstruction efficiency is measured from the unbiased sample of candidate second tracks.
A large effort was required by the study of 2-particle overlap in the detector, in order to reach the per mil accuracies.

Two kinematic fits to the $\mee\to X\gamma$ hypothesis (where $X$ allows for possible additional radiation) are performed for each event.
The parameters and covariance matrix of each charged-particle track, as well as the ISR photon direction are used in these fits.
The two-constraint (2C) `ISR' fit allows an undetected photon collinear with the collision axis.
The 3C `FSR' fit is performed only when an additional photon is detected. 
Most events have small $\chi^2$ values for both fits.
An event with only a small $\chi^2_{ISR}$ ($\chi^2_{FSR}$) indicates the presence of additional ISR (FSR) radiation.
Events where both fits have large $\chi^2$ values result from multi-hadronic background, track or ISR photon resolution effects, or the presence of additional radiated photons.
To accommodate the expected background levels, different criteria in the ($\chi^2_{ISR}$,$\chi^2_{FSR}$) plane are applied.
In the $\pi\pi\gamma$ channel, a loose 2D cut is used for the central $\rho$ region and a tighter cut for the $\rho$ tails. 
For $\rm KK\gamma$ the tight cut is used, while the region between tight and loose is exploited in efficiency studies.
The loose cut is also used in the $\mu\mu\gamma$ analysis. 
The $\pi\pi$, $\rm KK$ and $\mu\mu$ masses are calculated from the corresponding best `ISR' or `FSR' fit.

The evaluations of the acceptance and $\chi^2$ selection efficiency are sensitive to the description of radiative effects in the generator. 
The difference of the FSR rate between data and MC is measured and results in a small correction for the cross section. 
For additional ISR photons, more significant differences are found between data and the generator, since the latter uses a collinear approximation and an energy cut-off for very hard photons.  
The study of induced kinematical effects has been performed with the NLO {\small PHOKHARA} generator~\cite{phok} at four-vector level, with fast simulation. 
The differences occuring in acceptance yield corrections to the QED test. 
However, since radiation from the initial state is common to the pion, kaon and muon channels, the $\pi\pi(\gamma)$~($\rm KK(\gamma)$) cross section, obtained from the $\pi\pi$/$\mu\mu$~($\rm KK$/$\mu\mu$) ratio, is affected and corrected only at a few permil level.
The $\chi^2$ selection efficiency determined from muon data is applied to pions and kaons, after correcting the effect of secondary interactions, the $\pi/\mu$~($K/\mu$) difference for additional FSR, and kaon decays. 
The measured $\pi\pi(\gamma)$ and $\rm KK(\gamma)$ cross sections are almost insensitive to the description of NLO effects in the generator.

\section{The QED test} 

\begin{figure}[tb]
  \centering
  \includegraphics[width=8.0cm]{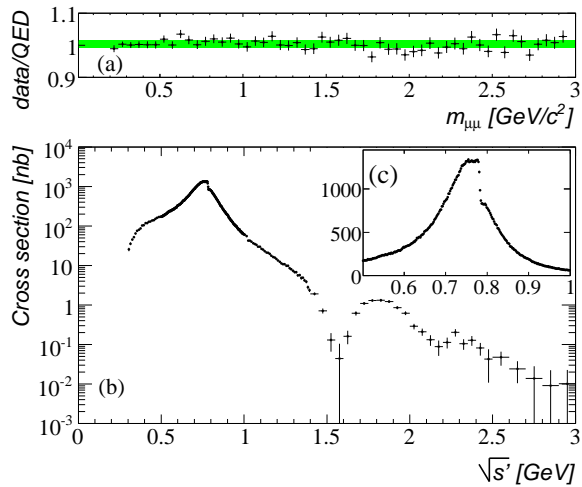}
  \caption{\small 
(a) The ratio of the measured cross section for 
$e^+e^-\to\mu^+\mu^-\gamma(\gamma)$ to the NLO QED prediction. 
The band represents a fit to a constant value~(see text). 
(b) The measured cross section for $e^+e^-\to\pi^+\pi^-(\gamma)$ 
from 0.3 to $3$\GeVE. 
(c) Enlarged view of the $\rho$ region in energy intervals of 2 MeV.
The plotted errors are from the sum of the diagonal elements of the 
statistical and systematic covariance matrices. }
\label{babar-log}
\end{figure}

The QED test involves two factors which cancel in the $\pi\pi$/$\mu\mu$~($\rm KK$/$\mu\mu$) ratio: $L_{ee}$ and the ISR photon efficiency, measured using a $\mu\mu\gamma$ sample selected only on the basis of the two muon tracks.
This test is expressed as the ratio of data to the simulated spectrum, after correcting for all known detector and reconstruction data-MC differences.
The generator is also corrected for its NLO deficiencies, using the comparison to {\small PHOKHARA}.
As shown in Fig.~\ref{babar-log}~(a), the ratio is consistent with unity from threshold to $3$\GeVM.
A fit to a constant value yields ($\chi^2/n_{\rm{df}}=55.4/54$)
\begin{equation}
\label{qed-test}
 \frac {\sigma_{\mu\mu\gamma(\gamma)}^{data}} {\sigma_{\mu\mu\gamma(\gamma)}^{NLO~QED}}~-~1~=~ (40\pm20\pm55\pm94)\times 10^{-4}~,
\end{equation}
where the uncertainties are statistical, systematic from this analysis, and 
systematic from $L_{ee}$ (measured using Bhabha scattering events), respectively.
The QED test is thus satisfied within an overall precision of 1.1\%.

\section{The $\pi\pi$ and $\rm KK$ cross sections}

An unfolding of the background-subtracted $m_{\pi\pi}$~($m_{\rm KK}$) distribution~(corrected for data/MC efficiency differences) is performed to correct for resolution and FSR effects.
A transfer matrix, obtained using simulation, provides the probability that an event generated in a given $\sqrt{s'}$ interval is reconstructed in a $m_{\pi\pi}$~($m_{\rm KK}$) interval.
The matrix is corrected to account for the larger fraction of events with bad $\chi^2$ values (and consequently poorer mass resolution) in data compared to MC, because of the approximate simulation of additional ISR.
The performance and robustness of the unfolding procedure have been assessed using data-driven test models~\cite{bogdan}.

Fig.~\ref{babar-log} (b, c) shows the results for the $e^+e^-\to\pi^+\pi^-(\gamma)$ bare cross section including FSR, $\sigma^0_{\pi\pi(\gamma)}(\sqrt{s'})$.
The main features are the dominant $\rho$ resonance, the abrupt drop at $0.78$\GeVE~ due to $\rho-\omega$ interference, a clear dip at $1.6$\GeVE~resulting from higher $\rho$ state interference, and some additional structure near $2.2$\GeVE.
The systematic uncertainties do not exceed statistical ones over the full spectrum, for the chosen energy intervals.
In particular, a systematic uncertainty of only $0.5\%$ has been achieved in the central $\rho$ region. 

A VDM fit of the pion form factor~\cite{bogdanSlides} was exploited to compare the BABAR data to other experiments. 
The BABAR data are described well by this fit in the region of interest for the comparison. 
There is a relatively good agreement~(within uncertainties) when comparing to the Novosibirsk data~\cite{cmd-2,snd} in the $\rho$ mass region, while a slope is observed when comparing to the KLOE '08 data~\cite{kloe08}. 
A flatter shape is observed when comparing to the more recent KLOE~\cite{kloe10,Babusci:2012rp} data, obtained by the analysis of events with a detected, large angle ISR photon. 
A good agreement is observed when comparing to the Novosibirsk and KLOE data, in the mass region below $0.5$\GeVM. 
There is a good agreement between the BABAR data and the most recent (isospin-breaking corrected)~$\tau$ data from Belle~\cite{belle}, while some systematic differences are observed when comparing to ALEPH~\cite{aleph05} and CLEO~\cite{cleo}.

\begin{figure}
\centering
\includegraphics[width=8.0cm]{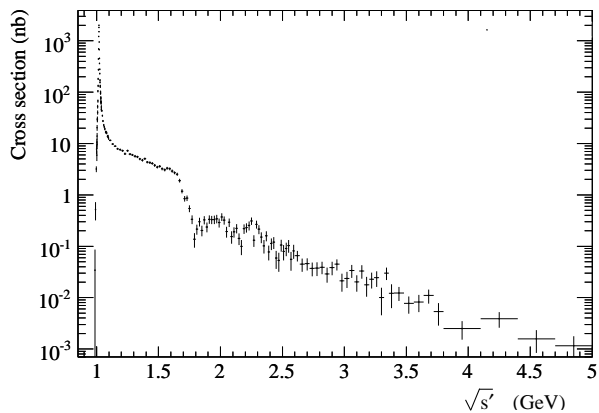}
\caption{\small  The measured $e^+e^-
    \to \KK(\gamma)$ bare cross  section (including FSR). Systematic and
    statistical uncertainties are shown, i.e., the diagonal  elements of the
    total covariance matrix. The contributions of the decays of the $J/\psi$ and
    $\psi(2S)$ resonances to $\KK$ have been subtracted.  
\label{Fig:Xsec_log}}
\end{figure}

The $\sigma_{\KK(\gamma)}^0(\sqrt{s'})$ cross section is shown in Fig.~\ref{Fig:Xsec_log}, from the $\KK$ production threshold up to $5$\GeVE.
The cross section spans more than six orders of magnitude.
Close to threshold it is dominated by the $\phi$ resonance, while other structures are clearly visible at higher masses. 
The contributions from the decays of the narrow $J/\psi$ and $\psi(2S)$ resonances to the $\KK$ final state have been subtracted for the cross section measurement and for the determination and parametrization of the kaon form-factor.
The systematic uncertainty in the $\phi$ region is of only $0.7\%$.

We fit the kaon form factor with a model~\cite{FFK-kuehn} based on a sum of resonances, for purposes of measuring the $\phi$ resonance parameters and providing an empirical parametrization of the form factor over the full range of the measurement.
The parametrized form factor is conveniently compared with the results of experiments at fixed energy values.
The fit is also necessary to extract the $\phi$ resonance parameters in the presence of other small contributions that need to be taken into account.
Since $\KK$ is not an eigenstate of isospin, both $\rm{I}=0$ and $\rm{I}=1$ resonances are considered.
Good agreement is found between the $\phi$ parameters obtained from the BABAR fit and the world average.

The measured charged kaon form factor is compared to data published by previous experiments~\cite{bogdanSlides}.
While the uncertainty of the BABAR cross section at the $\phi$ is $7.2\times 10^{-3}$, systematic normalization uncertainties of $2.2\%$ and $7.1\%$ are reported by CMD2 and SND, respectively.
The BABAR result, as well as the Novosibirsk measurements, are also affected by systematic uncertainties on mass calibration.
The observed mass differences are found to be compatible with the BABAR and CMD2 (SND) calibration uncertainties.
However, the normalization differences are not consistent by large factors with the quoted systematic uncertainties. 

The comparisons with the SND~\cite{SND:K}, OLYA~\cite{OLYA:K}, DM1~\cite{DM1:K}, and DM2~\cite{DM2:K} measurements is performed at higher masses too.
The systematic negative difference between BABAR and SND persists up to about $1.15$\GeVE, where a crossover occurs.
At higher masses, the SND values are consistently larger than the ones from BABAR.
The BABAR data are in rather good agreement with data from OLYA and DM1, while a systematic difference is obtained when comparing to DM2.

\section{The $\pi\pi$ and $\rm KK$ contributions to $a_\mu$} 

The lowest-order contribution of the $\pi\pi(\gamma)$ intermediate state to the muon magnetic anomaly is given by the integral
\begin{equation}
\label{eq:int_amu}
    a_\mu^{\pi\pi(\gamma),LO} \:=\: 
       \frac{1}{4\pi^3}\!\!
       \intl_{4m_\pi^2}^\infty\!\!\mathrm{d}s'\,K(s')\,\sigma^{0}_{\pi\pi(\gamma)}(s')~,
\end{equation}
where $K(s')$ is a known kernel~\cite{kernel}.
The integration uses the measured cross section and the computation of the uncertainties is done using the full statistical and systematic covariance matrices.
Each source of systematic uncertainties is taken to be fully correlated over the full mass range. 
The result of the integral from threshold to $1.8$\GeVE~ is
\begin{equation}
    a_\mu^{\pi\pi(\gamma),LO} \:=\: (514.1 \pm 2.2 \pm 3.1)\times 10^{-10}~,
\end{equation}
where the uncertainties are statistical and systematic.
This value is larger than that from a combination of previous $e^+e^-$ data ($503.5\pm3.5$), but is in good agreement with the updated value from $\tau$ decays ($515.2\pm3.4$)~\cite{newtau}.
When using the $\pi^+\pi^-$ data from BABAR only, the deviation between the BNL measurement~\cite{bnl} and the theoretical prediction is reduced to $2.4$ standard deviations.

The bare $e^+e^-\to\KK(\gamma)$ cross section obtained in the BABAR analysis is also used to compute the contribution of the $\KK$ mode to the theoretical prediction of the anomalous magnetic moment of the muon, following Eq.~\ref{eq:int_amu}. 
The result of the dispersion integral is 
\begin{equation}
\label{amukk-babar}
a_\mu^{KK,\rm LO}\!=\!\left(22.93\pm0.18_{\rm stat}\pm0.22_{\rm syst}\pm0.03_{\rm VP}\right)\times10^{-10},
\end{equation}
for the energy interval of interest, between the $\KK$ production threshold and $1.8$\GeVE.
The first uncertainty is statistical, the second is the experimental systematic, while the third is from the $\phi$ parameters used in the VP correction.
The precision achieved is $1.2\%$, the total error being dominated by the systematic uncertainties.
This is the most precise result for the $\KK$ channel, and the only one covering the full energy range of interest.
For comparison, the combination of all previous data~\cite{Davier:2010nc} for the same range yields $\left(21.63 \pm 0.27_{\rm stat} \pm 0.68_{\rm syst}\right)\times10^{-10}$.

\section{A fit to the BABAR $\rm K^+$ form factor in the high mass region} 

\begin{figure}
  \centering
  \includegraphics[width=8.0cm]{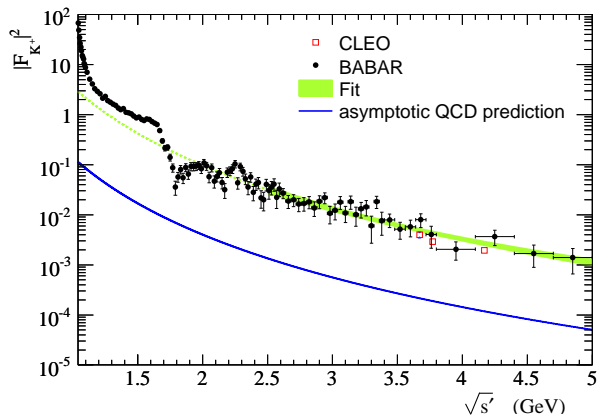}
  \caption{\small Fit~(green band)
    of the squared BABAR charged kaon form factor in the high mass
    region, using a function that has the shape of the QCD prediction~(blue
    curve, see text). The extrapolation of the fit at low energy is indicated 
    by the dotted green line.  We also indicate measurements from CLEO 
    data~(red squares), close to the $\psi(2S)$ mass and above.
    Systematic and statistical uncertainties are shown for data
    points~(i.e., the diagonal elements of the total covariance matrices).
  \label{Fig:FitFF_HM}}  
\end{figure}

At large masses~(i.e. above $2.5$\GeVM), the charged form factor can be compared to the asymptotic QCD prediction~\cite{chernyakFFK,BrodLepFFK}:
\begin{equation}
\label{eq:QCDasympt}
   F_K(s) = 16\pi \,\alpha_s\left(s\right) \, \frac{f^2_{K^+}}{s}. 
\end{equation}
The result of the fit of the squared form factor between $2.5$ and $5$\GeVE with the function $A \alpha_s^2(s) /s^n$~($A$ and $n$ being free parameters) is shown in Fig.~\ref{Fig:FitFF_HM}.
The contributions of the narrow $J/\psi$ and $\psi(2S)$ resonances decaying to $\KK$ are subtracted from the mass spectrum before performing the fit.

The fit describes the data well ~($\chi^2/n_{\rm{df}}=23.4/32$).
It yields $n = 2.04 \pm0.22$, which is in good agreement with the QCD prediction $n=2$.
The extrapolation of the fit to lower masses follows the average shape of the spectrum down to about $1.7$\GeVE.
However, the fitted form factor is about a factor of 4 larger than the perturbative QCD prediction of Eq.~(\ref{eq:QCDasympt}).
This confirms the normalization disagreement observed with the CLEO measurements~\cite{CLEOK,Seth:2012nn}, at masses near the $\psi(2S)$ and above.

\section{Conclusions and perspectives} 

BABAR has analyzed the $\pi^+\pi^-$, $\KK$ and $\mu^+\mu^-$ ISR processes in a consistent way, from threshold to $3(5)$\GeVM. 
The absolute $\mu^+\mu^-$ cross section has been compared to the NLO QED prediction, the two being in agreement within $1.1\%$. 
The $e^+e^- \to \pi^+\pi^- (\gamma)$~($e^+e^- \to \KK (\gamma)$) cross section, derived through the ratio of the $\pi^+\pi^-$~($\KK$) and $\mu^+\mu^-$ spectra is rather insensitive to the detailed description of radiation in MC. 
A strong point of the present analysis, comparing to previous ISR studies, comes from the fact that several uncertainties cancel in this ratio.
It allows us to achieve our precision goal: the systematic uncertainty in the central $\rho$ region ($0.6-0.9$\GeVM) is only $0.5\%$, and for the $\phi$ ($1.01-1.03$\GeVM) it is $0.7\%$.

The contribution to $a_\mu$ computed from the BABAR $\pi^+\pi^-$ spectrum, in the range $0.28-1.8$\GeVE, has a precision of $0.7\%$.
This is similar to the precision of the combined previous measurements. 
For the contribution to $a_\mu$ from the $\KK$ channel, the BABAR result is almost three times more precise compared to the previous world average.

In the comparison between the BABAR $\pi^+\pi^- (\gamma)$ cross section and the data from other experiments, there is a fair agreement with CMD2 and SND, while the agreement is poor when comparing with the various KLOE measurements. 
In order to make progress on this channel, the first priority should be to clarify the BABAR/KLOE discrepancy, the most important effect on $a_\mu$ being due to the difference on the $\rho$ peak. 
The origin of the slope in this comparison is also to be understood. 
The slope was very pronounced when comparing with the 2004 KLOE results, and it is reduced with the more recent KLOE data. 
The same slope is also observed in the comparison of the KLOE and $\tau$ data, while BABAR is in good agreement with the most recent $\tau$ results.

A fit of the charged kaon form factor has been performed using a sum of contributions from isoscalar and isovector vector mesons.
Besides the dominant $\phi$ resonance and small $\rho$ and $\omega$ contributions, several higher states are needed to reproduce the structures observed in the measured spectrum.
Precise results for the mass and width of the $\phi$ resonance have been determined, and are found to agree with the world average values.
In the $\phi$ region, discrepancies with CMD-2 and SND results are observed in the normalization of the cross section, the differences exceeding the uncertainties quoted by either experiment.
The results are in agreement with previous data at large energy, confirming also the large normalization disagreement with the asymptotic QCD expectation observed by the CLEO experiment.

\end{document}